\documentclass[
reprint,
 amsmath,amssymb,
 aps,
 pra,
floatfix
]{revtex4-2}

\usepackage{graphicx}
\usepackage{dcolumn}
\usepackage{bm}
\usepackage{amsmath}
\usepackage{amssymb}

\usepackage{xcolor, soul}
\sethlcolor{yellow}

\usepackage{textcomp}

\usepackage{hyperref}

\begin{document}

\preprint{APS/123-QED}
\title{Laser transfer and retrieval via nanophotonic supercontinuum process}

\author{Yongyuan Chu$^1$, Lu Yang$^1$, Wenle Weng$^2$, Junqiu Liu$^{3,4}$, Hairun Guo$^{1,4}$}
 
\email[Corresponding author: ]{hairun.guo@shu.edu.cn}
\affiliation{\looseness=-100
$^1$ Key Laboratory of Specialty Fiber Optics and Optical Access Networks, \\
Joint International Research Laboratory of Specialty Fiber Optics and Advanced Communication, \\ Shanghai University, Shanghai 200044, China \\
$^2$ Institute for Photonics and Advanced Sensing (IPAS), and School of Physics, Chemistry and Earth Sciences, \\
The University of Adelaide, Adelaide, South Australia 5005, Australia \\
$^3$ International Quantum Academy, Shenzhen 518048, China \\
$^4$ Hefei National Laboratory, University of Science and Technology of China, Hefei 230088, China
}%

\begin{abstract}
The nature of optical metrology is to perform efficient transfer and precise retrieval for lasers and optical signals, which is beneficial for a variety of applications ranging from optical clocking, spectroscopy, to telecommunications and quantum optics. 
While efforts have been made to promote the detection accuracy of optical frequencies, retrieval on optical waveforms remains on the autocorrelation scheme with limited performances. 
Here, we demonstrate a novel scheme for optical metrology, particularly on direct retrieval of optical waveform in terms of the field amplitude profile.
The scheme is based on massive four-wave-mixings underlying a nanophotonic supercontinuum process, which enables arbitrary transfer of an additive laser to modulational sidebands of the broadened continuum.
Detection of the transferred signals is then flexible to be within the whole span of the supercontinuum from visible to the mid-infrared range.
We demonstrate such a transfer scheme for both CW lasers and pulsed lasers. 
For the latter, the temporal amplitude profile of the optical wave can be retrieved, which reveals high-order dynamics of solitary pulses including the self-steepening, self-compression, and the soliton splitting, and shows a remarkable square-fold increase of signal-to-noise ratio in the power spectrum.
Our results may contribute to advance optical metrology particularly towards chip scale optical waveform detection, and more fundamentally, they reveal insights of massive ultrafast nonlinear interactions underlying the soliton-based supercontinuum process.
\end{abstract}

\maketitle

\section{Introduction}
\noindent
Optical metrology-measuring optical signals with high precision-has been of central importance in advancing science and applications ranging from optical clockwork \cite{diddams2001optical,herman2018femtosecond}, measure of physical quantities \cite{kuo1993force,graham2013new}, frequency synthesis \cite{cundiff2001optical,ma2004optical} and laser detection \cite{coddington2009rapid,fang2021vectorial}.
For the highest precision, an ultra-stable frequency reference linked to an ultra-stabilized atomic radiation is required, which is typically a microwave signal and provides an optical frequency standard by means of the microwave-to-optical conversion. Indeed, both the frequency chains of the early ages\cite{evenson1973accurate} and the revolutionary optical frequency comb technique \cite{telle1999carrier,udem2002optical,cundiff2003colloquium} have been developed as cornerstones for providing such a photonic-to-microwave link, in which optical nonlinear interactions were widely explored both for generating light at new frequencies and for coherent transfer of laser frequencies.

Of particular interest is the supercontinuum generation (SCG) process, which enables drastic optical spectral broadening to one or more octaves\cite{alfano1970emission}. 
Working with mode locked lasers that periodically emit a short intense pulse of laser and in the spectrum constitute a comb of exactly equidistant frequency elements, SCG could on one hand coherently expand the spectral coverage of the laser comb (i.e. adding more frequency elements on the side band while maintaining the same equidistance), and on the other hand enable the frequency stabilization by means of a self-referencing scheme \cite{holzwarth2000optical,jornod2018optical}. 
Therefore, such a coherent supercontinuum actually serves as an octave spanning optical frequency comb, and enables frequency metrology over a wide spectral range. 

Fundamentally, SCG is a rich bank of nonlinear optics, where a variety of nonlinear interactions work in a collective way in tailoring both the temporal and the spectral profile of the pulsed laser.
While SCG in fiber optics has been widely studied \cite{dudley2006supercontinuum,petersen2014mid}, increasing interest has been shifted to photonic integrated platforms where the efficiency of SCG can be largely boosted with increased wave confinement in sub-wavelength structures \cite{lamont2008supercontinuum,xia2021chip,oh2014supercontinuum}, and a broader range of nonlinear materials can be accessed providing a variety of emerging nonlinear interactions \cite{halir2012ultrabroadband,leo2014generation,johnson2015octave,liu2016octave,sinobad2018mid,yu2019coherent,kuyken2020octave,lu2020ultraviolet,may2021supercontinuum,fan2021higher,lee2023inverse,fan2024supercontinua}. This has led to new developments of coherent SCG. 
For example, the spectral broadening of photonic SCG is typically a few octaves under moderate pumping pulse energy, leading to a spectral coverage from the visible to the mid-infrared range \cite{singh2015midinfrared,yoon2017coherent,porcel2017two}. 
The dominant nonlinearity remains to be the Kerr effect from the cubic nonlinearity, yet quadratic nonlinear interactions are potentially a strong substitution, which could effectively lead to an engineered and enhanced Kerr-like nonlinearity\cite{moses2007self,zhou2012ultrafast,guo2014few}, and therefore drastically reduce the SCG threshold to few picojoules of the pulse energy \cite{hamrouni2024picojoule,jankowski2020ultrabroadband}. 
More significantly, simultaneous generation of the coherent supercontinuum and the higher order harmonics has been demonstrated in photonic waveguides \cite{okawachi2018carrier,hickstein2018quasi,lesko2021six}, such that optical beat between the laser and its harmonics, i.e. the self-referencing detection, is inline accomplished \cite{hickstein2017ultrabroadband,okawachi2020chip,carlson2017self,obrzud2021stable,ishizawa2022direct}. 
As such, photonic SCG can be beneficial for laser stabilization and optical frequency comb generation in high compactness\cite{obrzud2021stable}. 
Also, photonic SCG has been applied for spectroscopy\cite{nader2018versatile}, telecommunications \cite{hu2018single}, astronomic frequency calibration\cite{metcalf2019stellar}, frequency metrology\cite{carlson2017photonic}, and microwave synthesis. 
In particular, coherent SCG has provided an approach to access frequency combs in new wavebands including the mid-infrared range\cite{guo2018mid} and has enabled a mid-infrared dual-comb spectrometer for parallel gas phase detection in a sensitive, rapid, and accurate mode\cite{baumann2019dual,guo2020nanophotonic}.
More practically, it serves as a transfer oscillator that bridges both the target frequency and the referenced atomic radiation\cite{telle2002kerr,yao2021optical,yao2024coherent}.

\begin{figure*}[t]
\centering
\includegraphics[width=0.95\linewidth]{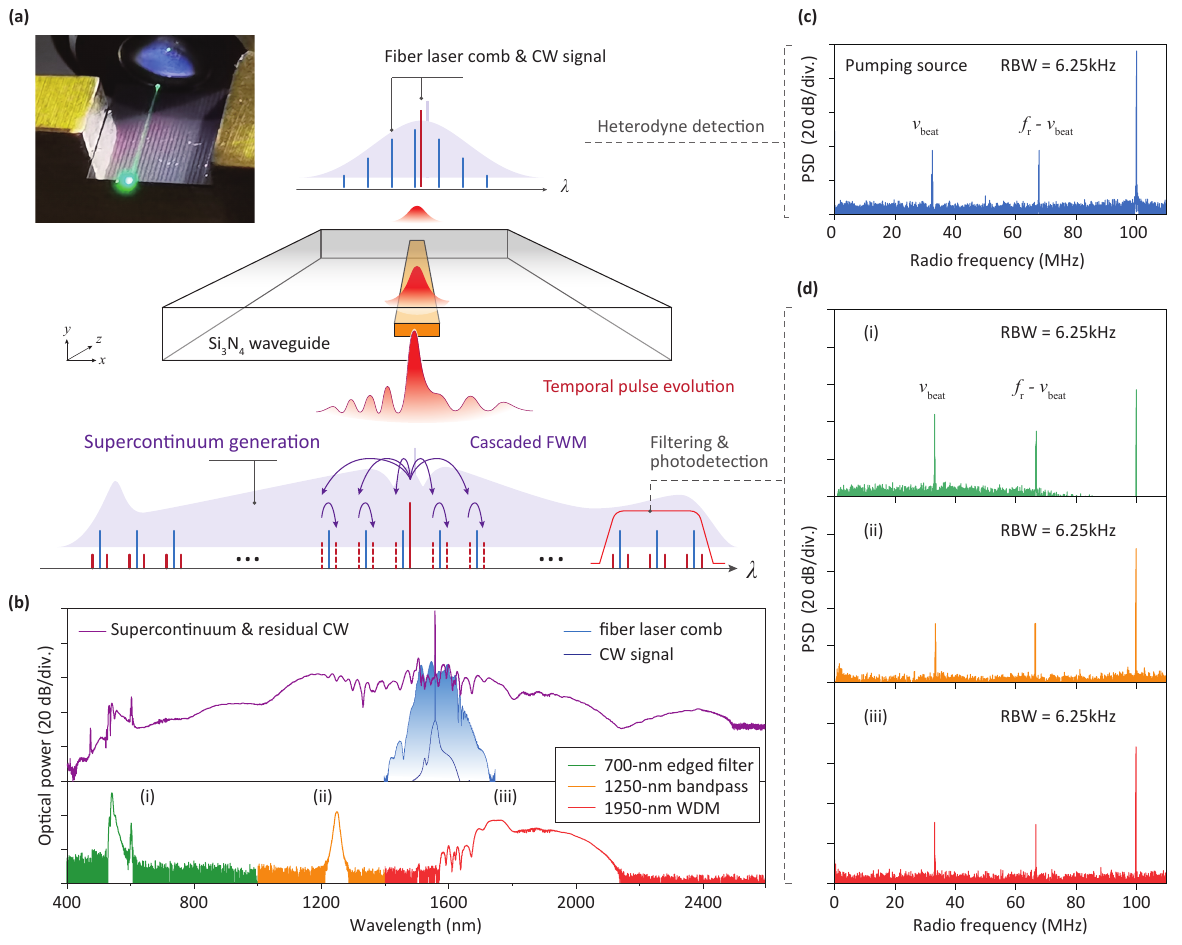}
\caption{
\label{fig_concept}
\textbf{Optical beat frequency transfer over the supercontinuum generation process.}
(a) Schematic diagram of the supercontinuum generation process in a photonic chip integrated and Kerr nonlinear waveguide, where a single frequency CW laser is co-launched with a fiber laser comb into the waveguide, and massive four wave mixing (FWM) processes are expected to occur during the light propagation. Inset is an experimental picture of the supercontinuum generation in a photonic $\rm Si_3N_4$ waveguide chip.
(b) A typical supercontinuum (purple curve) measured from a 5-mm $\rm Si_3N_4$ waveguide, together with the spectrum of the fiber laser comb (blue curve with shaded region) and the CW laser (dark blue curve) as the pumping source. The supercontinuum is further filtered and components in the visible region (marked as state (i), green curve), at 1250 nm (state (ii), yellow curve) and at the 1950-nm band (state (iii), red curve) are selected. 
(c) Direct heterodyne detection between the CW laser and the fiber laser comb, before the supercontinuum process.
(d) Measured \emph{rf} spectra underlying the selected optical spectral components of the supercontinuum. Each shows a pair of frequency tones similar to the heterodyne beatnotes of the pumping wave, which is understood by the CW laser transfer to modulational sidebands of the supercontinuum. The frequency tone $f_{\rm rep} = 100 ~{\rm MHz}$ indicates the repetition frequency of the fiber laser comb and of the supercontinuum as well.}
\end{figure*}

Optical heterodyne detection (also known as beat detection) is the follow-up operation that transfers optical signal to a microwave offset with respect to a referenced laser.
Conventionally, it is separated from the establishment of the optical frequency standard by nonlinear optical means (e.g. the coherent SCG).
However, merging the two processes would open up new opportunities for optical metrology.
For example, temporally downsampled sum-frequency generation (SFG) \cite{taschler2023asynchronous} and different frequency generation (DFG) \cite{bao2020interleaved} processes have been reported, allowing the generation of ultra-fine resolution frequency combs for spectroscopic applications, with detection signals simultaneously transferred to the radio frequency (\emph{rf}) domain.

In another panel, while developments have been focusing on increasing the detection accuracy of the laser frequency, measurement of the optical temporal waveform remains a challenge. 
Typically, tracing temporal pulse profile is done by interferometric detections (i.e. the autocorrelation measurement), by which the major pulse structure can be revealed, yet the access to ultrafast behaviors, especially the responses to high-order dispersive and nonlinear effects in the few-cycle regime, is hindered.
Frequency-resolved optical gating (FROG) \cite{trebino1993using} and spectral phase interferometry for direct electric field reconstruction (SPIDER) \cite{gallmann1999characterization} have been developed, allowing the retrieval of pulse spectral phase in addition to the intensity. 
Therefore, with full knowledge of the complex spectrum profile, the temporal electric field of the pulse is obtained.
However, FROG requires an iterative phase reconstruction algorithm to give a convincing guessed result, and SPIDER is complicated in configuring the spectral shearing interferometry. 
In this context, direct measurement of the temporal electric field for ultrafast pulsed lasers is still not allowed.

Here we demonstrate a novel approach for optical metrology by means of nonlinear optics underlying the supercontinuum generation in integrated photonics, which enables not only arbitrary frequency transfer over a wide wavelength span, but also the direct retrieval of optical temporal profile in terms of the field amplitude.
Physically, massive four-wave mixings are responsible for the frequency conversion of an additive laser inline with the SCG, leading to the generation of a pair of conjugate modulational sidebands underlying the supercontinuum.
This allows for laser frequency metrology at an arbitrary waveband of the continuum, e.g. in the visible, and for multi-laser metrology in a common band.
More importantly, in the transfer of pulsed lasers, the temporal profile of the optical field amplitude can be retrieved, whose power spectrum would feature a decent signal-to-noise ratio (SNR) of $\sim$60 dB, which is far better compared to the autocorrelation detection and on par with the state-of-the-art grating-based spectral analyzers.


\section{Results}
\noindent\textbf{Laser frequency transfer over SCG ---}
The concept of our work is illustrated in Fig. \ref{fig_concept}(a). 
When a laser frequency comb, temporally emitting an intense and ultrashort femtosecond pulse train, is launched into a Kerr nonlinear waveguide together with an additive CW laser, massive four-wave mixings (FWMs) are expected to occur within such a combined pumping wave over the propagation. 
Indeed, the pulsed pumping wave itself would experience self-phase modulation (SPM) from Kerr nonlinearity, which is the main cause for coherent SCG. 
In addition, the CW laser would via non-degenerate and cascaded FWMs be converted over the supercontinuum.
Fundamentally, the laser photons are transferred to both sides of frequency elements of the supercontinuum, forming a pair of conjugate sidebands (cf. Methods for related theoretical descriptions).
As such, by means of spectral filtering to select a fragment of the supercontinuum followed with photodetection, the conjugate sidebands are demodulated and shown as a \emph{rf} tone, whose frequency in nature would be identical to the heterodyne beat signal between the comb and the CW.

In an experiment to prove the above-mentioned laser transfer process (cf. Methods for experimental details), a fully stabilized femtosecond erbium-fiber laser (i.e. a laser comb) and a narrow-linewidth single frequency laser are combined and co-launched into a photonic chip based silicon nitride (${\rm Si_3N_4}$) waveguide.
To optimize the nonlinear coupling efficiency, both the CW and the laser comb are usually arranged in the same polarization direction (i.e. in the $p$-state), which further corresponds to one of the fundamental polarization directions of the $\rm Si_3N_4$ waveguide that has a close-to-rectangular cross section.
At the output of the waveguide, a typical supercontinuum spanning over two octaves was observed (cf. Fig. \ref{fig_concept}(b)), which is initiated by the laser comb and is independent of the presence of the CW (cf. Fig. \ref{fig_concept}(b)).
In particular, the supercontinuum would consist of dispersive waves both in the visible and in the mid-infrared regions\cite{guo2018mid}, determined by the overall dispersion landscape of the waveguide (cf. supplementary information (SI) for waveguide dispersion and engineered supercontinua).
The spectrum was further filtered, leaving a selected component for photodetection.
In detail, three separate spectral fragments, including a visible dispersive wave centered at $\sim 550 ~{\rm nm}$, are filtered and tested for laser frequency transfer.
As a result shown in the \emph{rf} spectrum, all components were found to have a pair of narrow linewidth signals (Fig. \ref{fig_concept}(d)), identical to the heterodyne beat signal with respect to the combined pumping wave (Fig. \ref{fig_concept}(c)). 
Therefore, the transfer of the CW laser is demonstrated, which is introduced in the initial pumping waveband and via SCG becomes detectable at an arbitrary waveband of the supercontinuum.
In addition, no degradation was found on the linewidth of such beat signals, which indicates a high coherence of the SCG process.

Practically, since clock transitions of several atoms (e.g. rubidium, cesium, potassium, strontium, and calcium ion) are typically in between $400-1000 ~{\rm nm}$ wavelengths, the demonstrated laser frequency transfer scheme allows for direct frequency stabilization (from a different wave band) with respect to such transitions, by means of the co-launched coherent SCG serving as the transfer oscillator.
In physics, given that the spectral coverage of the supercontinuum is on 100s THz and the repetition frequency ($f_{\rm rep}$) of the laser comb is on 100s MHz, the observed frequency transfer implies the existence of extensively massive and cascaded four-wave mixings up to the $10^6$-fold.

\begin{figure}[ht]
\includegraphics[width=1\linewidth]{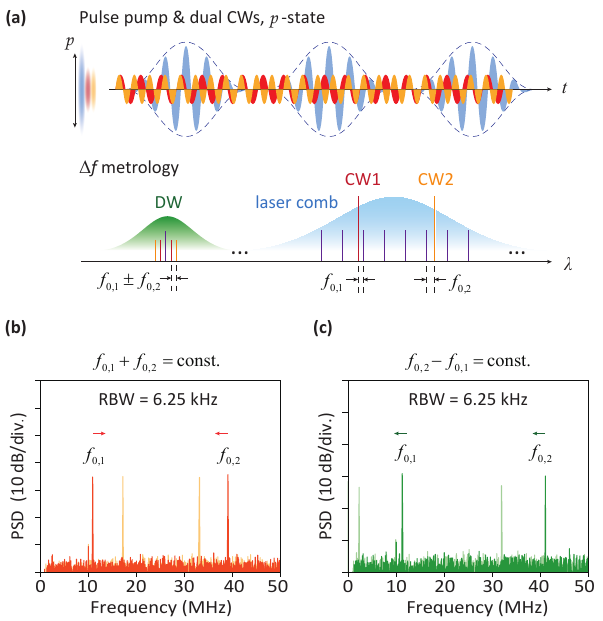}
\caption{\label{fig_metrology}
\textbf{Measuring laser frequency difference with transferred beatnote underlying the supercontinuum.}
(a) Schematic of $\Delta f$ metrology, in which the pumping wave consists of a laser frequency comb and two CWs, all in the same polarization direction (i.e. $p$-state).
Along with the supercontinuum process, CWs would be converted to the visible dispersive wave as two pairs of modulational sidebands.
(b, c) Measured \emph{rf} spectra underlying the visible dispersive wave of the supercontinuum, which reveals two frequency tones corresponding to $f_{0,1}$ and $f_{0,2}$. 
Therefore, the laser frequency difference is revealed by the knowledge of $f_{0,1}\pm f_{0,2}$. 
The light red trace in (b) and the light green one in (c) are individual shots after a slight shift of the CW lasers (with fixed offset), revealing ($+$) or ($-$) operation.
}
\end{figure}

\vspace{3 mm}
\noindent\textbf{Multi-laser transfer and ${\boldsymbol{\Delta f}}$ metrology ---}
Next, we investigated the transfer of multiple lasers along with the SCG process, which allows for precisely counting the frequency difference (${\Delta f}$) between lasers.
Typically, when two lasers (frequency $f_1$ and $f_2$) are far separated, measuring the frequency difference requires an optical frequency comb such that an optical-to-microwave link is established, i.e.:
\begin{equation}
    \Delta f = f_{0,1} \pm f_{0,2} + N \times f_{\rm rep}
\end{equation}
where $f_{0,1}$ and $f_{0,2}$ represent the offset frequency between the laser and the closest comb mode, which can be positive ($+$) or negative ($-$) valued, and $N$ should be a nonnegative integer.
The key to measuring $\Delta f$ with high precision is then to obtain the offset frequency value $f_{0,1} \pm f_{0,2}$.
Usually, $f_1$ and $f_2$ can be far away in frequency, located in different wave bands, so their detection should be operated individually. 
This is non-trival since a dedicated narrowband pass filter is always required to match the laser frequency, in order to suppress the residual comb modes and to leave an outstanding transfer beat.

However, following the laser frequency transfer scheme, the frequencies $f_1$ and $f_2$ can be effectively transferred over the SCG to a common waveband, converted as conjugate sidebands underlying the supercontinuum, which enables simultaneous detection of their offsets. 
As such, $f_{0,1} \pm f_{0,2}$ can be detected from an arbitrary wave band of the supercontinuum, especially in the visible region from the dispersive wave, where photodetectors usually show good performances and are cost-effective.
No narrow-band filtering is needed in this detection, yet a certainly wide waveband is essentially required for tracing the laser frequency, which corresponds to sufficient optical power such that the modulational sidebands can be resolved with a decent SNR.

The proof-of-concept experiment was then carried out and the results are shown in Fig. \ref{fig_metrology}.
In the experiment, a narrow-linewidth laser and an acoustic-optical modulator (AOM) were employed to produce two CW lasers with a controllable frequency offset.
When the two CWs and the laser comb are co-launched into the ${\rm Si_3N_4}$ waveguide (in the $p$-state), transfer of the lasers is expected along with the SCG process.
The visible dispersive wave packet of the supercontinuum was then filtered and photodetected, which exhibits two \emph{rf} tones corresponding to $f_{0,1}$ and $f_{0,2}$ (in the range up to $\frac{f_{\rm rep}}{2}$, cf. Fig.\ref{fig_metrology}(b)). 
In detail, the offset frequency was set as $\Delta f = 150 ~{\rm MHz}$ and $\Delta f = 130 ~{\rm MHz}$ as two states, and the repetition frequency of the laser comb is $f_{\rm rep} = 100 ~{\rm MHz}$.
Therefore, the observed \emph{rf} tones have $f_{0,1} + f_{0,2} = 50 ~{\rm MHz}$ and $f_{0,1} - f_{0,2} = 30 ~{\rm MHz}$, respectively.

The laser transfer scheme was also tested when the CWs and the laser comb are perpendicular to each other in the polarization direction (i.e. in the $s$-state and matching to the two fundamental polarized modes of the ${\rm Si_3N_4}$ waveguide), and was found successful as well (cf. SI for details).
This is mainly due to a non-standard cross section of the ${\rm Si_3N_4}$ waveguide that induces cross coupling in between the two fundamental polarized modes.
Also in the $s$-state, it is possible to isolate the residual CWs from the supercontinuum beam at the waveguide output.
As a consequence, we verify that the heterodyne-beat-like signal was in line with the supercontinuum, indicating a transfer in between the two polarization states in addition to the frequency conversion.

In practice, the proposed $\Delta f$-metrology can be useful, e.g. in the detection of high-frequency photonic microwaves from the gigahertz to terahertz range, and may relieve the requirements of ultrahigh-speed photodetectors as well as high-bandwidth signal analyzers.



\begin{figure*}[htbp]
\includegraphics[width=1\linewidth]{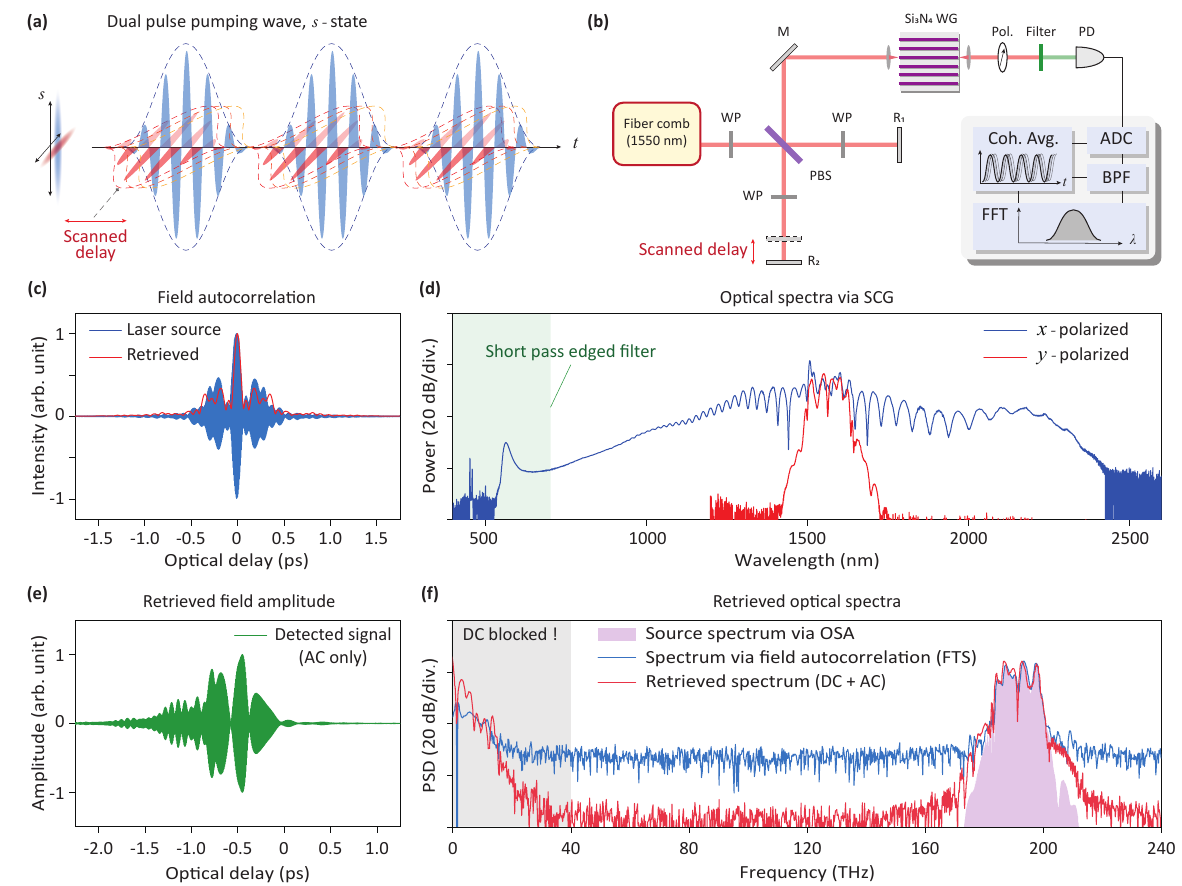}
\caption{\label{fig_pulse}
\textbf{Retrieval of temporal amplitude profile of optical pulsed wave via supercontinuum process.}
(a) Schematic of dual fiber laser combs crossing each other, which can be implemented either by using a scanned optical delay unit as illustrated, or by configuring dual combs with slight difference in the repetition frequency. The polarization state is controllable to be in the $s$- or $p$-state.
(b) Experimental setup for the retrieval of optical field amplitude profile (cf. Method for details). WP: wave plate, PBS: polarizing beam splitter, $R_{1,2}$: retroreflective mirror, M: reflective mirror, Pol.: polarizer, PD: photodetector, ADC: analog-to-digital converter, BPF: bandpass filter, Coh. Avg.: coherent averaging unit, FFT: fast Fourier transform unit.
(c) Directly measured autocorrelation trace of the pumping pulse (blue curve) and the trace from the retrieved field amplitude profile (red curve).
(d) Optical spectra at the output of the waveguide, where supercontinuum was observed in one polarization direction under the high pumping power, and the spectrum in the perpendicular polarization was maintained same as the pumping wave at low power.
(e) Retrieved temporal amplitude profile of the pumping wave, from the visible dispersive wave of the supercontinuum (the green shading range in (d)).
(f) Full power spectrum of the response signal underlying the visible dispersive wave (red curve), in which the AC component indicates the power spectrum of the pumping wave, which is compared with the FFT spectrum of the auto-correlation trace (blue curve) and with the direct measured spectrum using a grating-based commercial optical spectral analyzer (purple shaded curve).
}
\end{figure*}

\begin{figure*}[ht]
\includegraphics[width=0.82\linewidth]{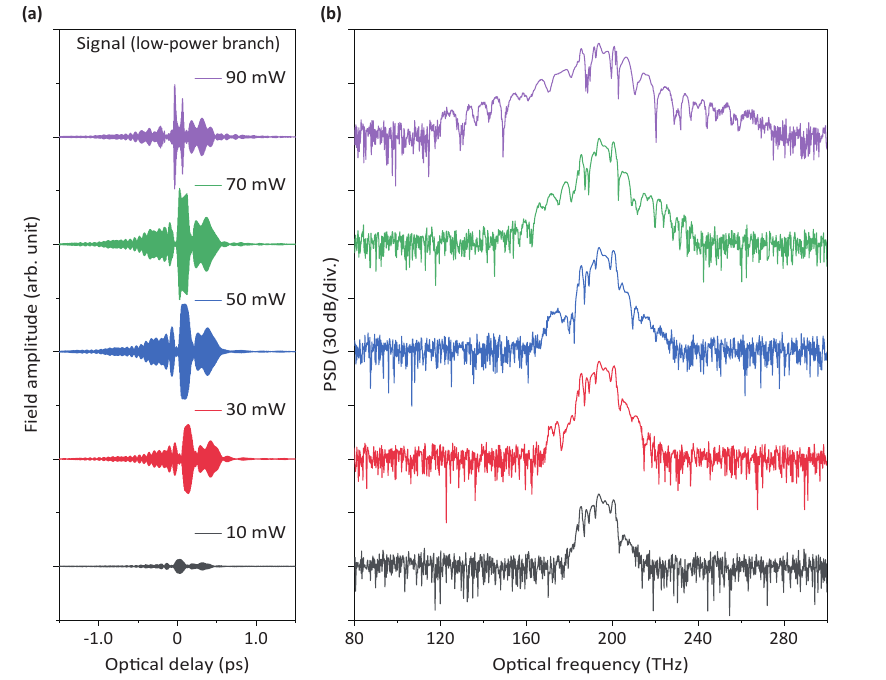}
\caption{\label{fig_evol}
\textbf{Evolution of temporal amplitude profile of pulsed wave with spectral broadening.}
(a) A set of retrieved field amplitude profiles underlying the visible dispersive wave, with a difference in the power of the pulsed waves in the low-power branch. The power of the high-power branch remains at $\sim ~60 {\rm mW}$ maintaining the supercontinuum.
(b) A set of power spectra corresponding to the field amplitude profiles in (a).
}
\end{figure*}

\vspace{3 mm}
\noindent\textbf{Pulsed laser transfer and retrieval of optical field amplitude profile ---}
Further, we investigated the laser transfer regime for pulsed lasers.
Fundamentally, a pulsed laser consists of multiple frequency elements that are equal-spacing and phase aligned (i.e. in the mode-locking state).
The key to transferring all the frequency elements with no cross-talk is that the repetition frequency of the targeted pulsed laser (i.e. the frequency spacing) should be different from that of the referenced comb, such that each element would be converted via SCG to a pair of conjugate side bands with different frequency offsets.
Therefore, the combined pumping wave (i.e. the targeted pulsed laser and the laser frequency comb) temporally acts as two asynchronous pulse trains crossing each other, which is similar to a dual-comb source with slightly different repetition frequencies. 
Physically, laser photons at each frequency become nondegenerate as conjugate photon pairs, and their recombination (by simply performing photodetection for the supercontinuum) would reveal the field amplitude profile as $\varepsilon_{\delta} + \varepsilon_{\delta}^{\dagger}$ (cf. Method for a detailed derivation), instead of the intensity ($|\varepsilon_{\delta}|^2$),
where subscription $\delta$ indicates the downconverted frequency of the photon, by comparing with the closest comb line of the reference frequency comb.
Therefore, the full amplitude of the pulse laser is ${\textstyle \sum_{i}^{}} {(\varepsilon_{\delta_i} + \varepsilon_{\delta_i}^{\dagger})}$, which is a \emph{rf} comb, and the frequency interval is determined by $\Delta f_{\rm rep}$.
Combining both the frequency downconversion and the direct retrieval of temporal field amplitude profile via Kerr nonlinearity, the proposed laser transfer regime is fundamentally distinguished from known techniques for the reconstruction of pulsed lasers.


In the experiment, we showcased the field amplitude detection for a femto-second fiber laser comb, see Fig. \ref{fig_pulse}.
To set up the asynchronous dual-comb pumping wave, Michelson interferometry was employed (cf. Fig. \ref{fig_pulse}(b)), in which the laser is split into two branches and via a scanned optical delay periodically crossing each other.
The fiber laser is power amplified so that one branch has sufficient power to support the supercontinuum generation in the $\rm Si_3N_4$ waveguide, while the other branch is in low power keeping the laser spectrum unchanged over the propagation.
Therefore, the fiber laser comb acts on the one hand as the referenced comb as well as the transfer oscillator, and on the other hand as the pulsed laser under test.
In fact, Michelson interferometry has been widely applied to obtain the field autocorrelation for a pulsed laser, from which the laser power spectrum can be retrieved.
The field autocorrelation of the pumping wave was recorded (Fig. \ref{fig_pulse}(c)), and the power spectrum was obtained as well (Fig. \ref{fig_pulse}(f)).
The pump beam was then launched into the $\rm Si_3N_4$ waveguide.
The polarization of both laser branches is managed in the $s$-state, matching the two fundamentally polarized modes of the waveguide.
At the output of the waveguide, the transmitted spectrum in each polarization (selected via a linear polarizer) was measured (Fig. \ref{fig_pulse}(d)).
A two-octave-spanning supercontinuum was observed from the high-power branch, while the spectrum in the low-power branch was almost unchanged. 
Next, the visible region of the supercontinuum was filtered and photodetected, which reveals a temporal response signal as a function of the optical delay time.
The one-way range of the optical delay is $\sim$50 ps. The detection resolution is as good as 0.1 fs (cf. Method for experimental details).
Therefore, the power spectrum of the response signal could effectively cover the full optical range of interest (physically up to a frequency of 100s THz).
In particular, the spectrum consists of both direct current (DC) and alternating current (AC) components (Fig. \ref{fig_pulse}(f)), in which the AC component is found almost identical to the power spectrum of the pumping wave.
Blocking the DC component, the temporal response signal then reveals the field amplitude profile of the wave (Fig. \ref{fig_pulse}(e)).
Convincingly, a field autocorrelation trace was also synthesized from the amplitude profile, which shows good agreement with the conventional measurement (cf. Fig. \ref{fig_pulse}(c)).

Technically, detecting and digitizing the amplitude signal would lead to a square-fold increase in the dynamical range of the laser power spectrum.
In our experiment, an increase of $\sim$30 dB in the spectral SNR (using a 10-bit ADC) was implemented, compared to the retrieved spectrum using the autocorrelation trace (cf. Fig. \ref{fig_pulse}(f)).
The overall dynamic range is as large as $\sim$60 dB, which is on par with the performance of grating-based optical spectrum analyzers.
More fundamentally, detailed structures of the temporal pulse are revealed from the field amplitude profile.
There is a central wave packet with highest amplitude, which is self-steepened in the ending edge. In addition, rich side lobes were observed underlying the pulse, which form the strong pedestal in the autocorrelation trace.

It is worth noting that the SCG and the laser frequency transfer are accumulatively occurring during the propagation in the $\rm Si_3N_4$ waveguide.
Therefore, the detected field amplitude profile and the retrieved spectrum represent an integrated result along the waveguide of a few millimeters, rather than a slice state at the input or the output.
This explains the slight difference between the retrieved spectrum and that of the initial pumping wave (cf. Fig. \ref{fig_pulse}(e,f)).
While the accuracy is also perturbed with unflat envelope of the supercontinuum, essential insights of ultrafast pulse dynamics, particularly the accurate measure of the primary wave packet, are preserved and revealed by the retrieval scheme.

Moreover, with an increase in the energy, pulse evolutional dynamics over the waveguide was characterized (cf. Fig. \ref{fig_evol}).
The field amplitude was first enlarged, followed by a split of its central waveform.
In the meantime, the split pulses were self-compressed in the duration and featured further enhancement in the peak amplitude (cf. Fig. \ref{fig_evol}(a)).
Consequently, the power spectrum is broadened and features oscillatory structures as a result of the pulse splitting (Fig. \ref{fig_evol}(b)).
Such an evolution matches with the universal understanding of pulse propagation in the soliton regime in Kerr nonlinear media with anomalous dispersion. 

\section{Conclusion}
In conclusion, we have proposed and investigated a laser transfer scheme that enables precise measurement for lasers in terms of both the frequency and the waveform.
The scheme is associated with coherent supercontinuum generation in a photonic integrated and Kerr nonlinear waveguide. 
In the co-propagation form, the laser photons are via four wave mixing processes converted as modulational and conjugate sidebands underlying the supercontinuum, and can be easily readout by performing photodetection for the continuum at an arbitrary waveband.
This enables not only robust laser frequency conversion within a wide range in the optical domain, but also the frequency transfer to the \emph{rf} domain for precise tracement.
More importantly, the scheme also allows for direct reconstruction of the temporal field amplitude profile for ultrafast pulsed lasers, by simple photodetection to recombine sideband conjugate photon pairs in the continuum.
Compared with conventional autocorrelation, measuring field amplitude leads to a square-fold increase in the dynamic range of the power spectrum. 
In practice, a signal-to-noise ratio of ca. 60 dB was implemented, which is competitive with state-of-the-art spectrometers.
As a consequence, essential dynamics of the laser pulses were unveiled from the field amplitude profile, including the soliton self-steepening, self-compression and the soliton splitting, which are popular in soliton based nonlinear propagation process, yet hindered by conventional detection methods.

Our work represents a distinguished approach to optical metrology, which may find applications in related areas such as laser detection and measurement, microwave photonics and spectroscopy. Fundamentally, by configuring a complex laser source to excite a nonlinear medium, our results contribute to the understanding of rich and high-order nonlinear interactions underlying the ultrashort soliton dynamics, which was not limited to the supercontinuum generation process, but in a broader range in ultrafast science.
Also, the use of chip integrated nonlinear waveguides shows a potential to revolutionize precise time and frequency measurements with high compactness.

\section*{Methods}
\noindent\textbf{Theoretical description of wave transfer to conjugate sidebands inline with the SCG:}
\newline We start with the generalized nonlinear wave equation that describes light beam propagation in a cubic nonlinear waveguide with dispersive effects, i.e.:
\begin{equation}
    \frac{{\partial \tilde A(\omega ,z)}}{{\partial z}} = 
    - i\beta (\omega )\tilde A - i\frac{\omega }{c}\frac{{{\chi ^{(3)}}}}{{2n{A_{\rm eff}}}}{\mathcal F}{\left[ {A^3(t,z)} \right]_\omega }
\label{eq_main}
\end{equation}
where $A(t,z)$ and $\tilde A(\omega ,z)$ indicate the propagation dynamic (along the $z$ axis) of the light field, in time ($t$) and in the physical (angular) frequency ($\omega $) domain, $\beta(\omega)$ indicates the wave propagation constant that contains the full dispersion of the waveguide, $\chi ^{(3)}$ is the nonlinear susceptibility of the waveguide material, $n$ is the effective refractive index of the waveguide, $A_{\rm eff}$ is the effective mode area that contains the transverse modal information of the light field, $c$ is the speed of light in vacuum.
The operator ${\mathcal F}$ indicates the Fourier transform, i.e. ${\mathcal F} \left[ A(t,z) \right]_\omega = \int dt {A(t,z)e^{-i \omega t}} = \tilde A(\omega ,z)$.
For the real-valued temporal field profile $A(t,z)$, the spectral profile $\tilde A(\omega ,z)$ features conjugate symmetry in the full frequency range $\omega \in (-\infty,\infty)$.

We further define the light field in the positive frequency range as a combination of a primary supercontinuum wave with dispersive wave, and a secondary CW for transfer, i.e.:
\begin{equation}
    {\tilde A}(\omega >0, z) = 
    ({\tilde A}_{\rm s}(\omega , z) + \tilde\sigma )e^{-i \beta _{\rm s} z} + {\tilde A}_c e^{-i \beta (\omega_{\rm c}) z} 
\label{eq_ans}
\end{equation}
where $\tilde A_{\rm s}$ indicates the supercontinuum wave packet in the soliton regime (soliton wave), which is centered at frequency $\omega_{\rm s}$ and has a dispersionless wave constant $\beta _{\rm s}(\omega) = \beta (\omega _{\rm s}) + (\omega-\omega_{\rm s}) \beta ^{(1)}(\omega _{\rm s}) + q$, here ${\beta ^{(m)}(\omega)}$ indicates the dispersion element of ${m}$-th order underline the wave constant profile, and $1/\beta ^{(1)}(\omega _{\rm s})$ is also known as the group velocity of the wave packet, $q$ is the nonlinear-induced phase shift.
The dispersive wave packet $\tilde\sigma$ is assumed to be on the edge of the supercontinuum, so its central frequency ($\omega_{\rm d}$) is away from that of the supercontinuum, i.e. $\omega_{\rm d} \neq \omega_{\rm s}$.
Being seeded with a laser comb, the supercontinuum inherits the comb structure to have a series of equal spacing frequency elements, that is, $\tilde A_{\rm s}(\omega) = \sum_{m} a_{\rm m}\delta(\omega_{\rm s}+2\pi m f_{\rm rep}) = \sum_{m} a_{\rm m}(\omega_{\rm m})$, and $a_{\rm m}$ indicates the amplitude of the comb element.
The CW is further expressed as: $\tilde A_{\rm c} = a_{\rm c}\delta(\omega_{\rm c})$, with constant amplitude $a_{\rm c}$ and frequency $\omega_{\rm c}$ that may have an offset compared to the comb, i.e. $\omega_{\rm c} = \omega_{\rm m} + \delta$.
The temporal profile of Eq.\ref{eq_ans} is marked as $A_{\rm s}(\tau, z) + \sigma(\tau, z) + A_{\rm c}(\tau)e^{i \phi z}$, which is in a co-traveling frame following the group velocity of the soliton wave, i.e. $t = \tau + \beta ^{(1)}(\omega _{\rm s})z$. 
The phase offset $\phi$ can be neglected if the CW frequency is close to the center of the soliton.

Substituting $\tilde A$ in Eq. \ref{eq_main} with Eq.\ref{eq_ans}, and by assuming that the evolution of the supercontinuum wave is slower than that of the dispersive wave (which is valid given a low soliton number underlying the SCG process), we have:
\begin{multline}
    \frac{\partial }{\partial z} \tilde\sigma (\omega ,z) = - i\Delta\beta(\omega)(\tilde\sigma + \tilde A_{\rm s}) \\ 
    -i\gamma{\mathcal F} {\left[ 2{\left| A_{\rm s} \right|}^2 \sigma + 2(A_{\rm s}A_{\rm c}^* + A_{\rm s}^*A_{\rm c})\sigma + \cdots  \right]}_{\omega > 0}
\end{multline}
where $\Delta\beta(\omega) = \beta(\omega)-\beta_s(\omega)$, and $\gamma = \frac{\omega }{c}\frac{{{3\chi ^{(3)}}}}{{8n{A_{\rm eff}}}}$. 
The first term on the right side of the equation describes the dispersive effect on dispersive wave generation.
Rapid growth of the dispersive wave is expected by having the phase match to the soliton wave, i.e. $\Delta\beta (\omega_{\rm d}) = 0$.  
The second term contains major nonlinear effects induced by the soliton and the CW, which includes not only the cross-phase modulation effect, but also the temporal modulation.
In fact, taking discrete elements underlying the soliton-based supercontinuum, a series of conjugate modulational sidebands can be derived, in which the first sidebands are always:
\begin{equation}
  a_{\rm m} a_{\rm c} e^{-i \delta \tau} + a_{\rm m} a_{\rm c} e^{i \delta \tau} = a_{\rm m}(\varepsilon_{\delta} + \varepsilon_{\delta}^{\dagger})
  \label{eq_notation}
\end{equation}
which reveal the beat between the CW and the closest element of the soliton wave. 
Moreover, given a flat spectral envelope of the supercontinuum (including the aligned phase), the amplitude of the frequency elements is likely to be identical, i.e. $a_{\rm m} \approx \bar a$.
Therefore, the amplitude of the CW is encoded in the modulation depth, which can be read out by simply performing photondetection for the dispersive wave packet or, more generally, for an arbitrary fraction of the supercontinuum outside the pumping range.
The CW frequency is down-converted, leaving a \emph{rf} offset ($\delta$) as the modulation frequency, which can be precisely detected with electronic devices.
This process also applies to pulsed lasers.
By simultaneously transferring all the laser frequency elements to modulational sidebands via the supercontinuum process, retrieval of the pulse waveform, i.e. ${\textstyle \sum_{i}^{}} {(\varepsilon_{\delta_i} + \varepsilon_{\delta_i}^{\dagger})}$, is enabled.

\vspace{3mm}
\noindent\textbf{Experimental setup and related parameters:}
\newline
In experiments, seeding supercontinuum generation uses an erbium-fiber-based femtosecond laser frequency comb (Menlo system Smartcomb) that emits temporal pulses with a duration $<70 ~{\rm fs}$. 
The comb is amplified in power. 
The pulse energy could reach $>1 ~{\rm nJ}$ while maintaining its short duration. 
The pulse repetition rate is $f_{\rm rep} = 100 ~{\rm MHz}$.
Such a pulsed laser is then launched into a photonic chip integrated $\rm Si_3N_4$ waveguide that usually has a close-to-rectangular cross section.
The $\rm Si_3N_4$ core is fully merged in the $\rm SiO_2$ cladding. The core height is typically $800 ~{\rm nm}$ and the width is lithographically engineered to tune the overall dispersion landscape of the waveguide. The length of the waveguide is $\sim 5~{\rm mm}$. An inverse taper structure is applied to the waveguide to increase the facet coupling efficiency, by which the total insertion loss of the waveguide is typ. $<3 ~{\rm dB}$.

As such, supercontinuum generation can be observed in both of the two fundamental polarized directions when pumping at the 1550-nm band (cf. SI for engineered SCG in $\rm Si_3N_4$ waveguide chips).
Physically, the supercontinuum process is governed by the soliton self-compression regime provided with anomalous group velocity dispersion of the waveguide at the pumping band.
Dispersive waves are enabled both in the visible and in the mid-infrared range, whose wavelengths are located in the normal dispersion range, further extending the spectral span of the supercontinuum.

The output supercontinuum could be further optimized to have a smooth envelope by tuning the pumping power or equivalently the seeding pulse energy.
The target is to shift the soliton first-compression point to the output facet of the waveguide, which is determined by the power-related nonlinear efficiency during the soliton propagation in the waveguide.
Also in this condition, a temporally compressed pulse train is obtained, which by proper dispersion engineering could be in the few-cycle state.

In our experiments, supercontinua with a smooth envelope are usually seeded at a pulse energy of $\sim 0.4 ~{\rm nJ}$. Further increasing the energy would access rich soliton dynamics after the first compression point, where the compression would relax and recur together with the soliton splitting under high-order dispersive and nonlinear perturbations. The spectrum would then feature an oscillatory envelope as the result of temporal interference of split solitons.

In the transfer of CW lasers, a narrow-linewidth tunable laser is co-lined with the laser comb and co-launched into the waveguide. 
The linewidth of the laser is typ. $<10 ~{\rm kHz}$.
The required laser power is a few tens of milliwatts, leading to a more than 30 dB SNR in the transfer beats.

In the transfer and retrieval of the pulsed laser, the complex dual comb source is configured using a Michelson interferometer (cf. Fig. \ref{fig_pulse}(b)).
In both pathways, beam polarization is well managed, which in corporate with a polarized beam splitter would allow for free tuning of the laser power.
The recombined beam is in the $s$-state.
Reading out the field amplitude is by photodetection of the visible dispersive wave of the supercontinuum supported by the high-power branch.
In the $s$-state, a linear polarizer is first used to remove the transmitted light from the low-power branch and select the supercontinuum beam merely, followed by a spectral bandpass filter to choose the visible dispersive wave component, which has a power of a few milliwatts.

The photodetection leads to a temporal response trace as a function of the optical delay time. The overall delay range is 50 ps, by means of a shaking mirror at a frequency of 1 Hz. Therefore, the one-way detection time is 0.5 s. 
The bandwidth of the photodetector and the sampling rate of the data acquisition unit are both above 1 MHz, which corresponds to an optical resolution of 0.1 fs.
Hence, the observation bandwidth is sufficiently large to cover the optical band of interest (e.g. the 1550-nm band for the laser retrieval).
Moreover, coherent averaging was applied to increase the SNR of the response signal and to probe the dynamic range of the acquisition system. 
The normalized SNR is estimated to be $10 /{\sqrt s}$ (cf. SI for details).

The power spectrum of the response is then obtained by the Fourier transform of the temporal trace, in which the AC component appears at a similar frequency as the pumping wave and shows an almost identical spectral envelope.
The DC component is induced by the nonlinear cross-coupling in between the pumping waveform, indicating an additive perturbation from the low-power pulse to the high-power supercontinuum when they are temporally overlapped (cf. SI for details). 
Blocking the DC component, the temporal response trace would become symmetrical up and down with respect to the zero-level, which is recognized as the field amplitude profile of the wave.


\vspace{3mm}
\noindent\textbf{Acknowledgments:} 
We acknowledge funding from National Key Research and Development Project of China (No. 2020YFA0309400), National Natural Science Foundation of China (No. 11974234, 12374313), and Innovation Program for Quantum Science and Technology (2023ZD0301500). This work is also supported by 111 Project (D20031) by MoE of China. Silicon nitride waveguide chips were foundry fabricated by LIGENTEC and by Qaleido Photonics.
We also acknowledge fruitful discussions with Prof. Longsheng Ma and Prof. Yanyi Jiang at ENUC, which makes us aware that in parallel to our work, the single laser transfer was also implemented in the fiber-based supercontinuum process, enabling a coherent link crossing different wavebands (Ref.\cite{yao2021optical,yao2024coherent}). Some of our earlier results, including the photonic chip based supercontinuum dynamics probed with the dual-comb pumping waves (termed as vector supercontinuum process) and the single laser transfer behavior, were presented at Conference on Lasers and Electro-Optics Europe (CLEO Europe) 2021 (ee\_p\_6), Nonlinear Optics 2021 (NTu2A.3), and CLEO 2022 (JW3B.48).

\vspace{3mm}
\noindent\textbf{Author Contributions:}
Y.C. and H.G. conceived the experiments. H.G. and J.L. designed the photonic integrated silicon nitride waveguide chip. Y.C. and L.Y. characterized the waveguide chip and performed the experiments under the supervision of H.G.. Y.C., L.Y. W.W. and H.G. analysed the experimental data. Y.C. and H.G. wrote the manuscript with contributions from all others. H.G. supervised the whole project.

\vspace{3mm}
\noindent\textbf{Conflict of interest:} 
Y. Chu, L. Yang, and H. Guo are inventors of a patent application related to this work. Others declare no conflicts of interest.

\vspace{3mm}
\noindent\textbf{Data Availability Statement:} 
Source data and simulation codes related to this work will be released on the repository Zenodo upon publication of this preprint.

\nocite{*}
\bibliography{main}

\clearpage
\newpage
\appendix

\renewcommand{\thefigure}{S\arabic{figure}}
\setcounter{figure}{0}
\renewcommand{\theequation}{S\arabic{equation}}
\setcounter{equation}{0}

\section*{S\lowercase{upplementary Information}}
\begin{figure*}[ht]
\centering
\includegraphics[width=1\linewidth]{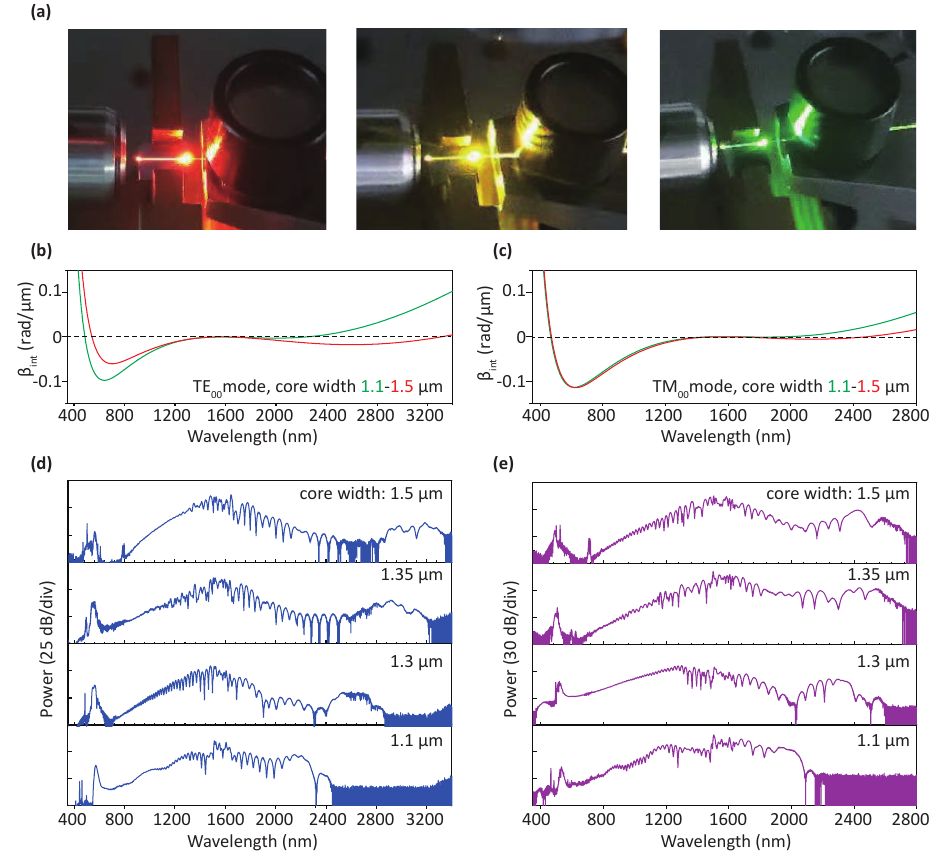}
\caption{
\label{fig_s1}
\textbf{Nanophotonic supercontinuum generation in $\rm{Si_3N_4}$ waveguide chips.}
(a) Pictures of on chip supercontinuum generation with different visible colors from the dispersive wave. (b,c) Strip of integrated dispersion profile in the waveguide in two fundamentally polarized modes, in which the boundary profiles correspond to the waveguide core width $1.1 ~{\rm \mu m}$ and $1.5 ~{\rm \mu m}$. (d,e) Set of Supercontinua in waveguides of different core width, and in both the $\rm TE_{00}$ and the $\rm TM_{00}$ mode. The central wavelength of the pumping wave is ca. 1560 nm. The wavelength of dispersive waves are predicted by the dispersion profile, and are tuned with a change of the waveguide core width. The waveguide core height is ca. 870 nm, and the length of the waveguide is ca. 5 mm. The total insertion loss of the waveguide is ca. 3 dB. The pumping pulse energy is ca. 0.6 nJ at the input facet.}
\end{figure*}
\section*{D\lowercase{ispersion engineered supercontinuum generation in nanophotonic waveguide chips}}
\noindent Nanophotonic supercontinuum generation was based on $\rm Si_3N_4$ waveguide chips that feature wide transparency, high Kerr nonlinearity, and high flexibility for dispersion engineering. Indeed, dispersion engineering is implemented by simply changing the geometric cross section size of the waveguide, or more specifically the core width, and therefore shares the lithographic precision of nanofabrication in tailoring the waveguide size\cite{guo2018mid}. 
Typically, the $\rm Si_3N_4$ waveguide core is fully claded by $\rm SiO_2$, and has a core width of $700\sim900 ~{\rm nm}$ to support anomalous group velocity dispersion (GVD) for the soliton-based supercontinuum process. The core width can be flexibly changed from a few hundred nanometers to a few micrometers. As such, the full dispersion of the waveguide is tuned. The waveguide usually features two zero-GVD points in wavelength, which indicates the generation of dispersive waves in the normal GVD region both in the visible and in the mid-infrared, when pumping in the near-infrared (typ. at the 1550-nm band) governed by the anomalous GVD. In this way, engineering the waveguide dispersion is mostly reflected as the wavelength change of the dispersive waves. Physcially, dispersive wave generation is by its phase matching to the soliton that has a dispersion-less phase profile, i.e.:
\begin{align}
\Delta \beta(\omega) &= \beta(\omega) - \beta_{\rm s}(\omega) \nonumber \\
&= \beta(\omega) - \Bigl[ \beta(\omega_0) + (\omega - \omega_0)\beta^{(1)}(\omega_0) \nonumber \\
&\quad + \frac{1}{2}(\omega - \omega_0)^2\beta^{(2)}(\omega_0) + q \Bigr]
\end{align}
where $\beta(\omega)$ indicates the wave propagation constant in the waveguide that contains the full dispersion information, $\beta^{\rm (1)}(\omega)$ is the first-order derivative indicating the wave group velocity, $\omega_0$ is the central frequency of the soliton wave, and $q$ is the soliton-induced nonlinear phase shift that is usually small valued and ignorable. Therefore, the phase difference between the soliton and dispersive waves is mostly induced by all order of dispersion in an integral form, i.e.: 
\begin{equation}
\beta_{\rm int}(\omega) = \sum_{m\ge2}{(\omega-\omega_0)^m \beta^{\rm (m)}(\omega_0)/m!} \approx \Delta \beta(\omega)
\end{equation}

Here, we present a set of supercontinua in $\rm Si_3N_4$ waveguides with a variation in the core width, which feature the wavelength change of dispersive waves (cf. Fig.\ref{fig_s1}). The pumping wave is an erbium-fiber-based femtosecond fiber laser comb, with a pulse duration of $<70 ~{\rm fs}$, and the pulse energy amplified to be $>1 ~{\rm nJ}$. In each waveguide, supercontinuum is supported in both of the two fundamentally polarized modes, namely the $x$-polarized $\rm TE_{00}$ mode and the $y$-polarized $\rm TM_{00}$ mode. The dispersion profile of all waveguide modes was calculated using the finite element method (COMSOL), which successfully predicts the wavelength of the dispersive waves. Usually, the mid-infrared dispersive wave is more sensitive to structural change compared to that in the visible region. But still, dispersive waves of different visible colors were implemented. 
Such results lay a solid foundation for this work exploring higher-order nonlinear dynamics excited by a complex pumping wave.
\begin{figure*}[ht]
\includegraphics[width=1\linewidth]{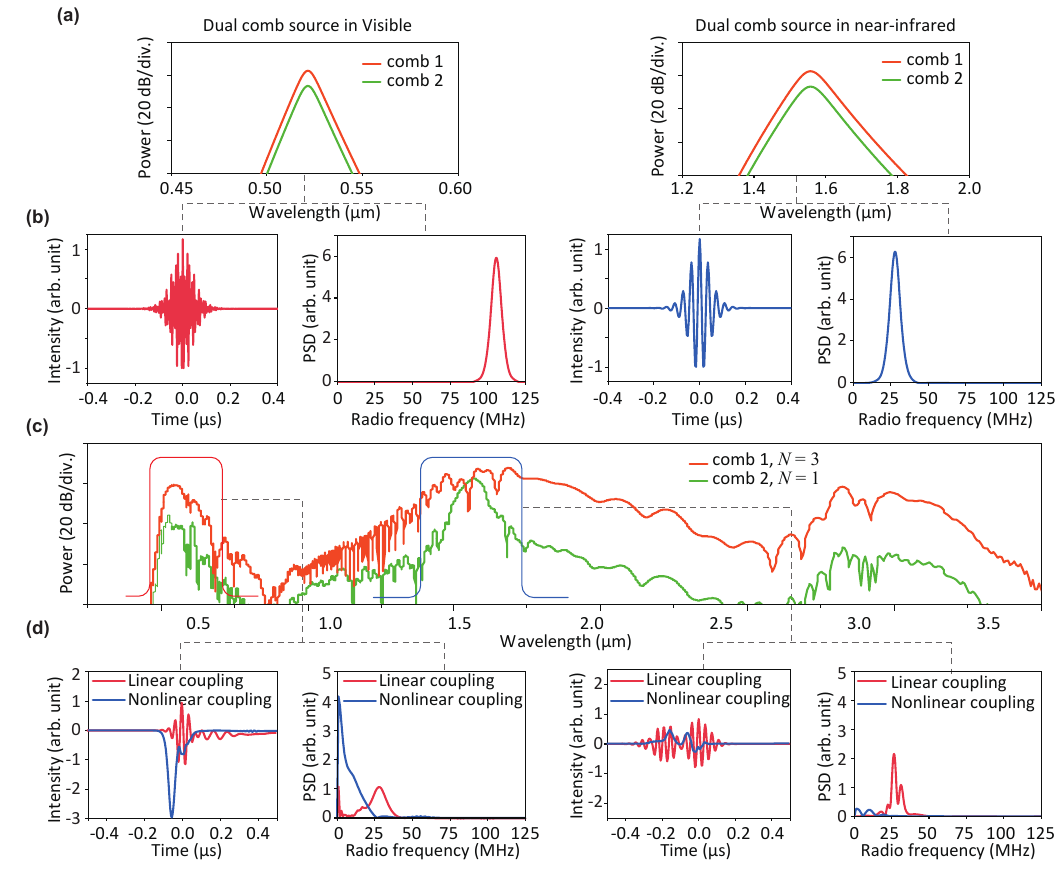}
\caption{
\textbf{Numerical simulation probing the laser transfer response underlying the supercontinuum generation upon dual-comb pumping wave.}
(a) Exampled dual comb laser sources at the lasing wavelength $\sim525~{\rm nm}$ and $\sim1550~{\rm nm}$. (b) The field autocorrelation and the corresponding radio spectrum of the dual-comb source. (c) Generated spectra upon dual comb propagation in the waveguide modes with the coupling effects, when the delay in between is $\Delta t = 0$. The high-power comb (comb 1, orange curve) is evolved to a supercontinuum spanning over three octaves, while the low-power comb (comb 2, green curve) maintains its central spectrum. (d) Radio frequency response extracted from the supercontinuum wave (comb 1), namely from the visible dispersive wave and from the central portion (1550-nm band). The response is further separated into the DC and AC components, upon the control of parameters $\kappa$ and $\gamma_{\rm XPM}$. Note: for both the field autocorrelation traces and the autocorrelation-like response traces, the base line (i.e.the zero-value) is corresponding to the averaged laser power when two pulsed waves are far separate in the time domain. 
}
\label{fig_s2}
\end{figure*}

\begin{figure*}[ht]
\includegraphics[width=1\linewidth]{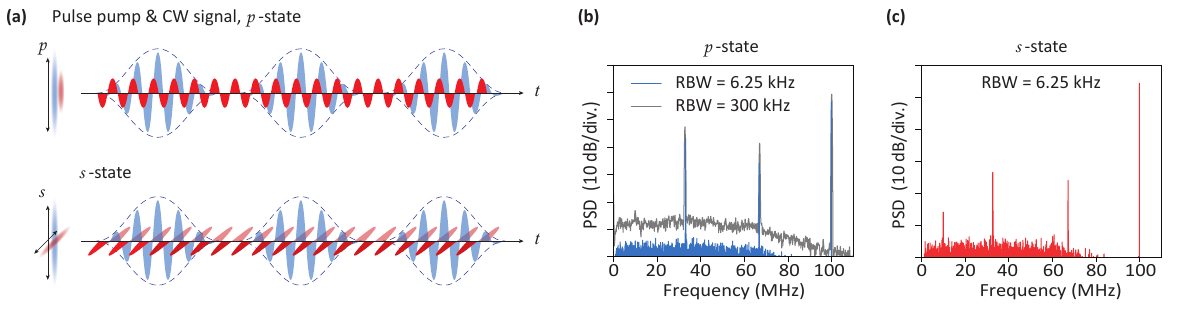}
\caption{
\textbf{Single CW laser frequency transfer in $p$- and $s$-states.}
(a) Schematic diagrams for laser beams in the $p$- and $s$-states. (b) Transferred beat signal in both states, and the SNR of the beat is optimized by tuning the power of the initial CW laser. The signal is extracted by photodetecting the visible dispersive wave of the supercontinuum. The RBW of the radio frequency spectrum is 6.25 kHz and 300 kHz.
}
\label{fig_s3}
\end{figure*}

\section*{S\lowercase{imulation of supercontinuum and laser transfer response upon dual-comb pumping wave}}
The full model for the assessment of the photonic supercontinuum process and the underlying response upon the dual comb pumping wave is based on a group of coupled nonlinear Schrödinger equations (NLSEs), i.e.:
\begin{align}
\frac{\partial \tilde{A}_1(\Omega, z)}{\partial z} 
    &= -\frac{\alpha_1}{2} \tilde{A}_1 - i \beta_{1,\mathrm{int}} \tilde{A}_1 - i\kappa \tilde{A}_2 \nonumber \\
    &\quad -i\gamma_{1,\mathrm{SPM}} \mathcal{F}\left[ \left\lvert A_1 \right\rvert^2 A_1(\tau, z) \right]_{\Omega} \nonumber \\ 
    &\quad -i\gamma_{\mathrm{XPM}} \mathcal{F}\left[ \left\lvert A_2 \right\rvert^2 A_1(\tau, z) \right]
\\
\frac{\partial \tilde{A}_2(\Omega, z)}{\partial z} 
    &= -\frac{\alpha_2}{2} \tilde{A}_1 -i\beta_{2,\text{int}} \tilde{A}_2 -i\kappa \tilde{A}_1 \nonumber \\
    &\quad-i\gamma_{2,\mathrm{SPM}} \mathcal{F}\left[ \left\lvert A_2 \right\rvert^2 A_2(\tau,z) \right]_{\Omega} \nonumber \\ 
    &\quad -i\gamma_{\mathrm{XPM}} \mathcal{F}\left[ \left\lvert A_1 \right\rvert^2 A_2(\tau,z) \right]
\label{eq_s1}
\end{align}
where ${\tilde A}_{j}(\Omega,z)$ is the spectral envelope of the two pulsed waves ($j=1,2$) upon propagation in two of the waveguide supported modes, and is expressed in a relative frequency frame $\Omega = \omega-\omega_0$, $A_{j}(\tau,z)$ is the corresponding temporal amplitude profile defined in a co-traveling time frame ($\tau = t - \beta_{1}^{\rm (1)}(\omega_0)z $) following the group velocity of the primary wave ($j=1$), $\beta_{j,\rm int}(\Omega) = \beta_j(\Omega+\omega_0) - (\beta_1(\omega_0) + \Omega\beta_{1}^{\rm (1)}(\omega_0))$ indicates the dispersion landscape of the two waveguide modes, $\kappa$ is the linear coupling coefficient between the two waves, $\gamma_{j,\rm SPM} \propto \frac{n_2}{A_{j,\rm eff}}$ is the nonlinear coefficient of the self-phase modulation (SPM) process ($n_2$ is the nonlinear refractive index and $A_{j,\rm eff}$ is the SPM-related effective mode area), and $\gamma_{\rm XPM}$ is that of the cross-phase modulation (XPM) process. For waves in the two fundamentally polarized modes and with a similar effective mode area, one may conclude that $\gamma_{\rm XPM} = \frac{2}{3}\gamma_{j,\rm SPM}$. The Raman effects were not included in this model, nor the higher-order nonlinearities. The operator $\mathcal F[~]_{\Omega}$ indicates the Fourier transform operation that connects the time domain ($\tau$) and the frequency ($\Omega$) domain. 

The initial state of the two seeding waves further reads $A_{1}(\tau,0) = a_{1}(\tau,0)$ and $A_{2}(\tau,0)=a_{2}(\tau+\Delta t,0)e^{i{\omega_0}{\Delta t}}$, where $\Delta t$ indicates a time delay between the two waveforms. 
Therefore, the simulation of waves' propagation would be repeated over the delay time, modeling consecutive pulse pairs from the dual-comb source\cite{coddington2016dual} (or from a Michelson interferometer\cite{kwiat1990correlated}) as the pumping wave.
Upon interference of two pulsed waves, the temporal interferogram (as a function of $\Delta t$, also known as the field autocorrelation trace\cite{legendre1993spatial}) would then show the carrier wave oscillation determined by $e^{i{\omega_0}{\Delta t}}$.
For dual combs with a slight difference in the repetition frequency ($\Delta f_r$), the carrier frequency is further down converted to a radio frequency $\Delta\omega = \Delta f_0 + n\Delta f_r$ ($\Delta f_0$ is the difference of the carrier envelope offset frequency, $n$ is an integer), and the interpulse delay is discrete as $\Delta t = m\frac{\Delta f_r}{f_{r,1}f_{r,2}}$ ($f_{r,j}$ is the repetition frequency of the comb wave, $m$ is an integer). Therefore, we have $A_{2}(\tau,0)=a_{2}(\tau+m\frac{\Delta f_r}{f_{r,1}f_{r,2}},0)e^{i{\Delta \omega}\frac{m}{f_{r,1}}}$. 
In this context, the carrier frequency ($\Delta \omega$) of the dual-comb interferogram would vary over the lasing wavelength (linked to $n$).
In another panel, the laser transfer regime would also lead to an autocorrelation-like response signal, yet revealing a distinguished feature with the interferometric response, namely the carrier frequency would be \emph{unchanged} throughout the supercontinuum.
In fact, this crucial distinction was qualitatively verified by numerical simulations based on the coupled NLSEs model (cf. Fig. \ref{fig_s2}).

In the simulation, the repetition frequency of dual combs was $f_{r}\sim 250 ~{\rm MHz}$, with $\Delta f_r \sim 320 ~{\rm Hz}$ and $\Delta f_0 \sim 25 ~{\rm MHz}$. This led to a field autocorrelation centered at the radio frequency of $\sim 25 ~{\rm MHz}$ for the pumping dual combs at the 1550-nm band, and that for the visible lasing waves (e.g. at the wavelength of $\sim 525 ~{\rm nm}$) was at $\sim 100 ~{\rm MHz}$; see Fig. \ref{fig_s2}(a,b)).
When the dual-comb waves are launched into the waveguide modes, they would be in the soliton regime supported by the anomalous GVD of the waveguide, and via soliton self-compression to have a broadened spectrum.
The soliton number $N$\cite{demircan2007analysis}is defined to scale the level of the compression and that of the spectral broadening in a standard condition (i.e. with only GVD and the SPM effect), which is:
\begin{equation}
N^2 = \frac{L_D}{L_N}=\frac{T_0^2\cdot\gamma_{j,\rm SPM} P_0}{|\beta^{(2)}(\omega_0)|}
\end{equation}
where $L_D$ and $L_N$ are the so-called dispersion length and the nonlinear length that scale the dispersive and nonlinear effects on the soliton propagation, $T_0$ indicates the pulse duration of the comb wave and $P_0$ is the pulse peak power. 

By tuning $P_0$, one of the comb waves was set to have $N=3$ for the supercontinuum generation, and the other is $N=1$ for \emph{no} spectral broadening.
The spectrum of both waves after 5-mm propagation is then simulated (cf. Fig.\ref{fig_s2}(c)).
The high-power wave evolved to a supercontinuum with a span of almost three octaves, and features dispersive waves in both the visible and the mid-infrared region.
The central spectrum of the low-power wave is almost unchanged, yet by intermode coupling it has a weak copy of the supercontinuum.
Such a two-wave simulation was repeated over the delay time (or for a number of $m$), and in each one, the power of the visible dispersive wave from the supercontinuum wave is recorded, forming the autocorrelation-like response signal as well as the response spectrum (cf. Fig. \ref{fig_s2}(d)).
Similar to experimental observations, there are both DC and AC components in the response signal.
Moreover, the AC component is found connected with the linear coupling ($\kappa \neq 0$) between the two waves, and the DC component is related to the nonlinear coupling effects ($\gamma_{\rm XPM} \neq 0$).
Interestingly, extracted from the visible wave, the carrier frequency underlying the response was still $\sim 25 ~{\rm MHz}$.
This implies that the characteristics of the initial pumping wave (encoded as the interferometric signal) have been transferred to new wavebands by the supercontinuum process.
Physically, the response signal by itself indicates the power fluctuation of the supercontinuum, which is the result of the pulse collision\cite{roy2005dynamics} of two comb waves upon the propagation in the waveguide.
Therefore, the proposed numerical model could qualitatively confirm our experimental observations, particularly the transfer regime for pulsed lasers.

\vspace{3mm}
\section*{E\lowercase{ffects of }CW\lowercase{ laser frequency transfer in different polarization state}}
\noindent Here, we present more experimental results to give a full view of the laser transfer regime.
Usually, the photonic waveguides support two fundamentally polarized modes, so there are options to configure the polarization states for both the pulse-seeding supercontinuum beam and the laser beam under test.
Clearly, the optimized solution is to manage the two beams in the same polarization state, i.e. in the $p$-state, and further align them to one of the fundamental modes of the waveguide. As such, the modal overlap between the beams is maximal, and same for the efficiency of the laser transfer.
Moreover, given a non-perfect waveguide cross section, coupling between the two fundamental modes is also allowed but with a much reduced efficiency.
Thus, the two beams can be perpendicular in the polarization direction, i.e. in the $s$-state, and by being aligned to both of the two fundamental modes in the waveguide feature modal coupling as well as the laser transfer.
In another panel, the power of the laser is an essential factor that can impact the signal-to-noise ratio (SNR)\cite{johnson2006signal} of the transferred beat signal (or the response signal).
\begin{figure*}[!th]
\includegraphics[width=1\linewidth]{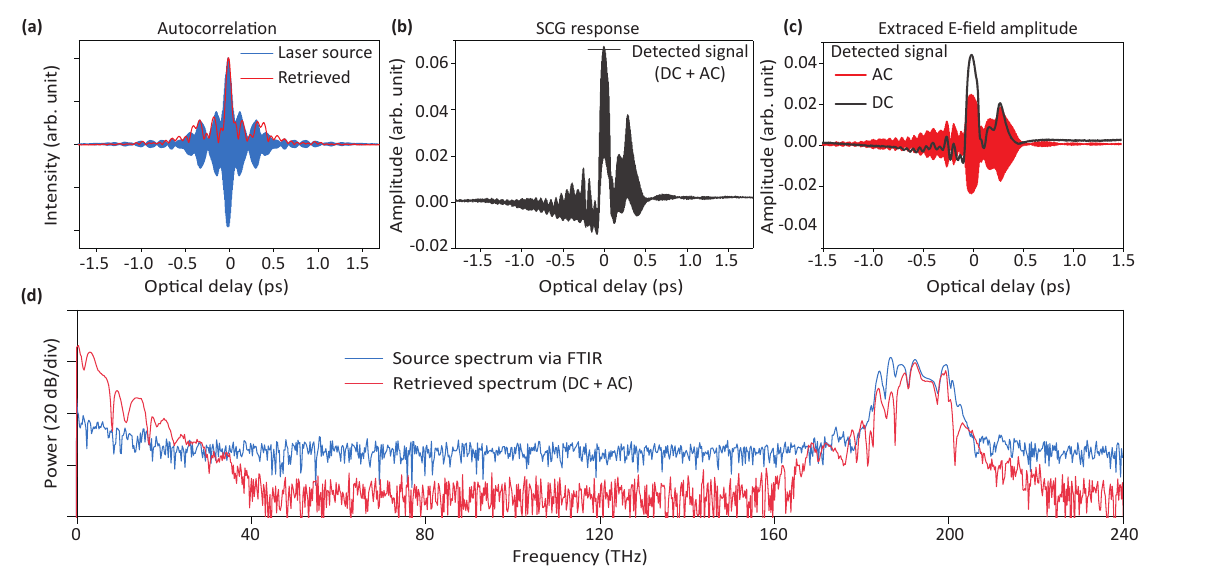}
\caption{
\textbf{Pulsed laser retrieval in the $p$-state polarization.}
(a) Direct measured field autocorrelation on the combined pumping source (blue curve) and the synthesized trace from the retrieved optical field amplitude profile (red curve). (b) Full autocorrelation-like response signal extracted from the visible dispersive wave of the supercontinuum (which was first spectrally filtered followed by the photondetection). (c) DC and AC traces from the full response signal, in which the AC trance reveals the field amplitude profile. (d) Fourier spectrum (power spectrum) of the full response signal (red curve) and the source spectrum measured by the FTS (blue curve). 
}
\label{fig_s4}
\end{figure*}
We tested the laser frequency transfer for a CW laser, in both the $p$- and $s$-states (cf. Fig.\ref{fig_s3}), and compared their performance.
In each case, the transferred beat signal is measured from the visible dispersive wave of the supercontinuum, and the SNR of the beat is optimized by tuning the CW laser power in between the $10\sim100 ~{\rm mW}$. 
While the SNR in the $s$-state could barely reach $\sim 30 ~{\rm dB}$, that in the $p$-state is $\sim 50 ~{\rm dB}$ benefit from the maximal modal overlap.
Therefore, we demonstrate that the laser transfer scheme is effectively enabled in both the $s$- and $p$-states, particularly the latter is much preferred and feasible for the frequency locking operations.

\section*{P\lowercase{ulsed laser transfer in the $p$-state}}
\noindent We further testified the transfer and the retrieval of the pulsed laser when it was combined with the supercontinuum beam in the same polarization state (i.e. the $p$-state) in the waveguide (cf. Fig.\ref{fig_s4}).
In this configuration, the high power branch to support the supercontinuum is ca. 50 mW, and the low power branch is ca. 5 mW, both beams are coupled to the fundamental TE mode of the $\rm Si_3N_4$ waveguide.
Again, the visible dispersive wave from the supercontinuum is the selected waveform that was filtered to characterize the autocorrelation-like response as well as the field amplitude profile.
The full autocorrelation-like response signal (cf. Fig.\ref{fig_s4}(b)) shows strong asymmetry in the amplitude with respect to the referenced zero-level (which indicates the average power of the visible dispersive wave when two pulsed beams are temporally far separated).
The DC and AC components of the response were further extracted (cf. Fig.\ref{fig_s4}(c)), from the latter a field-autocorrelation trace is synthesized and is consistent with the direct measured autocorrelation on the pumping source (cf. Fig.\ref{fig_s4}(a)).
This confirms again that the AC component of the response reveals the field amplitude profile of the pulsed laser.
From the Fourier spectrum of the response, the AC component in the spectrum was also found close to that of the source spectrum measured by the Fourier transformed spectrometer\cite{griffiths1983fourier} (cf. Fig.\ref{fig_s4}(d)), yet has certain deviation. 
The improvement on the dynamical range was also demonstrated using the supercontinuum supported laser transfer scheme, which has a ca. 20-dB lower noise background and more than 50-dB spectral SNR.

\section*{E\lowercase{stimation on the signal to noise ratio of the response signal}}
\begin{figure*}[ht]
\includegraphics[width=1\linewidth]{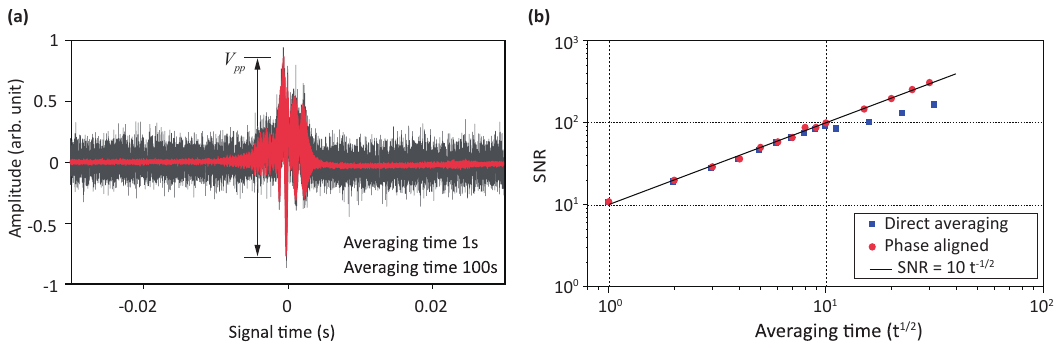}
\caption{
\textbf{Estimation on the SNR of the response signal}
(a) Response signals for pulsed laser transfer and retrieval, over the averaging time of 1 s (i.e. the single-shot trace, in gray) and 100 s (in red), respectively. Averaging would lead to the reduction of the noise on the background. (b) Statistic analysis on the normalized SNR, which by coherent averaging shows the trend of ca. $10 ~t^{-0.5}$ (red points). As a comparison, direct averaging would also lead to an increase on the SNR, which follows the same trend within the detection time of 100 s, yet shows a deviation over a longer time.
}
\label{fig_s5}
\end{figure*}
\noindent Basically, the response signal extracted from a fraction of the supercontinuum to the dual-pulse pumping source is a periodic trace over the reversible time delay. 
The signal may feature fluctuations period to period due to potential jittering of the laser or instability of the mechanical delayline, which is reflected as temporal phase drift.
The SNR of the single-shot trace is mainly determined by the efficiency and the coherence of the nonlinear wave-mixing processes underlying the supercontinuum, which transfer the pulsed laser from the pumping waveband to the dispersive wave region on edge.
In this context, the noise of the signal is mostly white Gaussian noise \cite{djuric1996model}(cf. Eq.\ref{eq_s6} as the distribution of the noise amplitudes, where $\sigma$ indicates the standard deviation of the distribution), which can be reduced by averaging such that the overall SNR can be improved.
The ultimate SNR is then purely limited by the dynamical range of the digital acquisition unit in measuring the signal.
However, while the level of the white noise can be reduced by the averaging, the signal with phase fluctuations may also be degraded and the overall SNR could not reach the ultimate level.
To overcome the signal fluctuation, re-alignment of the phase on the signal is possible, which is followed by the averaging and could raise the SNR in a standard exponential trend. This process on the data is termed coherent averaging\cite{guo2020nanophotonic}.
\begin{equation}
f\left ( x \right ) = \frac{1}{\sqrt{2\pi } } e^-{\frac{\left ( x-\sigma  \right )^{2} } {2\sigma ^{2} } }
\label{eq_s6}
\end{equation}
\begin{equation}
{\rm SNR} = \log_{10}{\left ( \frac{V_{pp}}{2\sigma }  \right ) } 
\label{eq_s7}
\end{equation}

Here, we probed and estimated the ultimate SNR by means of the coherent averaging.
As the result shown in Fig. \ref{fig_s5}, the peak SNR of the response signal was considered (cf. Eq. \ref{eq_s7})
The detection time for each period of the signal is 1 Hz (determined by the mechanical optical delay line).
The initial SNR corresponding to a single period of the response was ca. 10, which can be increased by increasing the detection time (in which the number of signals for averaging is increased).
It was clear that by coherent averaging, the increase of the SNR follows the trend of ca. $10 ~t^{-0.5}$, namely the normalized SNR for the signal.
As a comparison, directly averaging the signal (without phase re-alignment) would follow the same trend within the detection time of 100 s, yet shows a deviation over a longer time.

\end{document}